\documentclass[%
 reprint,
 nofootinbib,
 amsmath,amssymb,
 aps,
]{revtex4-2}

\usepackage{graphicx}
\usepackage{dcolumn}
\usepackage{bm}
\usepackage{hyperref}
\usepackage{siunitx}
\usepackage{comment}

\begin{document}


\title{A First Proof Of Principle Booster Setup\\For The MADMAX Dielectric Haloscope}
\hfill MPP-2019-259

\author{Jacob Egge}
\author{Stefan Knirck}%
 \email{knirck@mpp.mpg.de}
\author{Béla Majorovits}
\author{Christopher Moore}
\author{Olaf Reimann}
\affiliation{%
 Max-Planck-Institute for Physics\\
 Föhringer Ring 6, 80805 Munich, Germany
}%

\date{\today}

\begin{abstract}
Axions and axion-like particles are excellent low-mass dark matter candidates. The MADMAX experiment aims to directly detect galactic axions with masses between $40\,\mu{\rm eV}$ and $400\,\mu{\rm eV}$ by using 
the axion-induced emission of electromagnetic waves from  boundaries between materials of different dielectric constants under a strong magnetic field.
Combining many such surfaces, this emission can be significantly enhanced (boosted) using constructive interference and resonances.
We present a first proof of principle realization of such a booster system consisting of a copper mirror and up to five sapphire disks. The electromagnetic response of the system is investigated by reflectivity measurements.
The mechanical accuracy, calibration process of unwanted reflections and the repeatability of a basic tuning algorithm to place the disks are investigated. 
We find that for the presented cases the electromagnetic response in terms of the group delay predicted by one-dimensional calculations is sufficiently realized in our setup. The repeatability of the tuning is at the percent level, and would have small impact on the sensitivity of such a booster.
\end{abstract}


\maketitle


	\section{Introduction}
	Many astrophysical observations indicate the existence of cold dark matter (CDM) -- a non-radiating and weakly interacting matter also present in our galaxy~\cite{Bertone:2004pz}. 
	The axion is an excellent candidate to make up CDM~\cite{DINE1983137,PRESKILL1983127,ABBOTT1983133,DAVIS1986225,LYTH1992189,Kawasaki:2014sqa,Fleury:2015aca,Ringwald:2015dsf,Fleury:2016xrz,Borsanyietal,Ballesteros:2016xej}, while it has been originally proposed to solve the strong CP-problem~\cite{PhysRevLett.38.1440,PhysRevLett.40.223,PhysRevLett.40.279}.
	The cosmological production of the observed abundance of axion dark matter depends on whether the Peccei-Quinn (PQ) symmetry is broken before or after cosmic inflation.
	Matching this abundance with the observed dark matter abundance gives a prediction for the axion mass.
	While in a pre-inflationary scenario the axion mass $m_a$ is relatively unconstrained below ${m_a \lesssim \SI{e-2}{\electronvolt}}$~\cite{PhysRevD.98.030001}, the post-inflationary symmetry breaking scenario restricts it to roughly $\SI{20}{\micro\electronvolt} \lesssim m_a \lesssim \SI{200}{\micro\electronvolt}$~\cite{PhysRevD.85.105020,Kawasaki:2014sqa,Fleury:2015aca,Klaer:2017ond,Gorghetto:2018myk,Kawasaki:2018bzv}.
	
	Several experimental efforts are underway to directly probe galactic axions; for a review cf.~\cite{Graham:2015ouw,Irastorza:2018dyq,PhysRevD.98.030001}. Most approaches rely on the conversion of axions to photons under a strong magnetic field, which can be resonantly enhanced by a cavity~\cite{Sikivie:PhysRevLett.51.1415,Sikivie:PhysRevD.32.2988}. Since dark matter moves non-relativistically, the frequency $\nu$ of the converted photons is set by the axion mass up to small velocity corrections $\sim 10^{-6} \nu$.
	Most notably, the ADMX collaboration has excluded QCD axion models for most masses between $\sim \SI{2.7}{\micro\electronvolt}$ and $\sim \SI{3.3}{\micro\electronvolt}$ with this approach~\cite{Braine:2019fqb}, but also other experiments such as HAYSTAC~\cite{Zhong:2018rsr} provide some of the most sensitive limits.
	At higher masses one needs to reduce the volume of resonant cavities to still match the wavelength of the converted photons. In addition, higher frequency cavities typically have lower quality factors. Both effects decrease sensitivity of cavity experiments for increasing~$m_a$.
	Hence, a variety of different concepts to generalize the cavity setup to use higher modes in larger volumes have been proposed and are being prototyped, such as cavities with multiple coupled cells, e.g.\ pizza cavities~\cite{Jeong:2017hqs} or RADES~\cite{Melcon:2018dba}, cavities with dielectrics, e.g.\ \cite{McAllister:2017ern,Kim:2019asb}, or simply the idea to scan in parallel with multiple separate~cavities,~e.g.~ORGAN~\cite{McAllister:2017lkb}.
	
	An alternative approach to cavity setups is the dish antenna~\cite{Horns:2012jf}. Placing a metal mirror inside a strong magnetic field, CDM axions can convert to electrogmagnetic radiation on its surface. This generates an emitted power of $\sim \SI{e-27}{\watt\per\square\meter}$ for a magnetic field of \SI{10}{\tesla}, independent of the axion mass. While this allows for a broadband measurement at high masses, the emitted power is still not enough to reach competitive sensitivies with realistic setups.
\begin{figure}
	\centerline{
		\includegraphics[width=\linewidth]{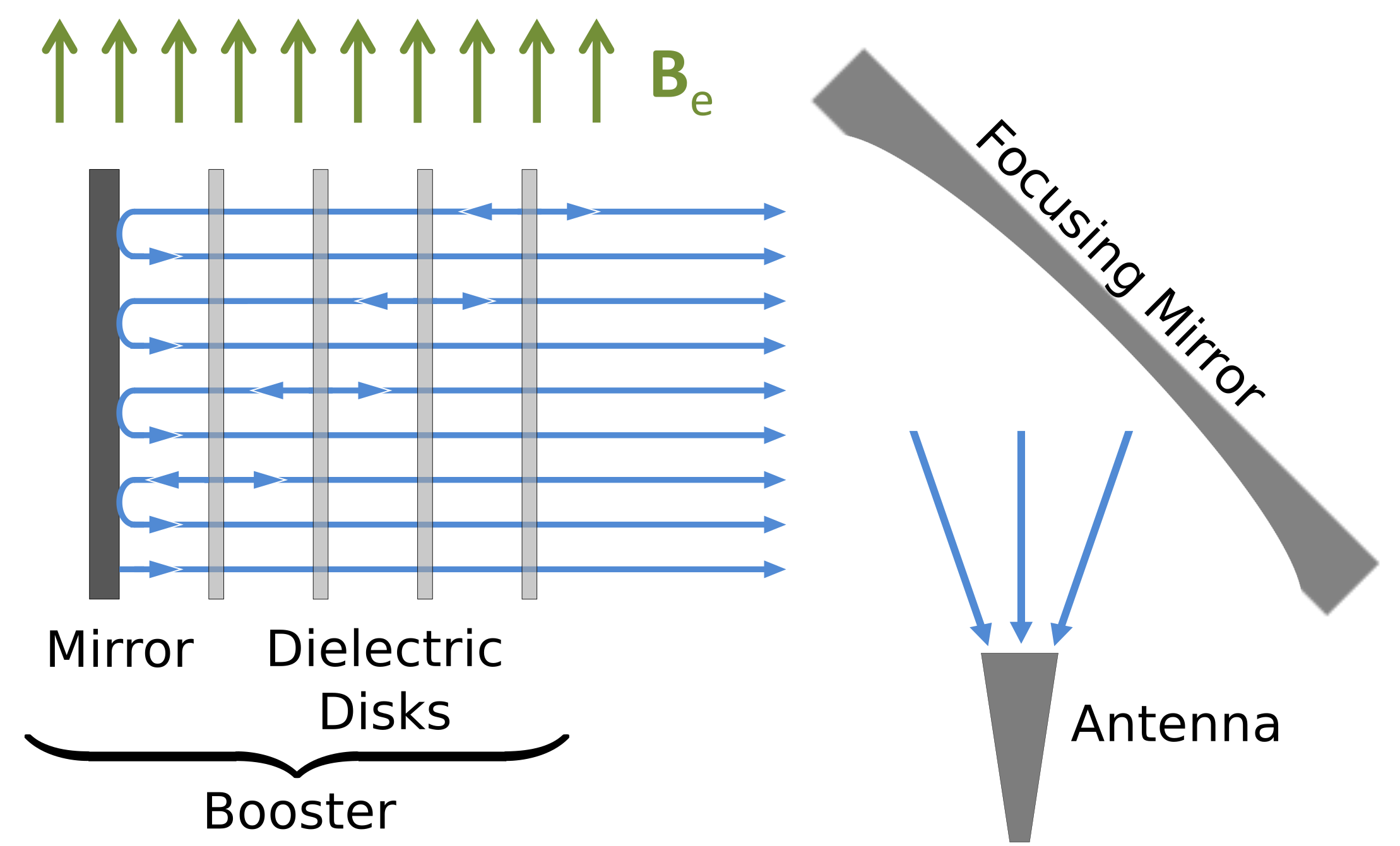}
	}
	\caption{
		Conceptual sketch of a dielectric haloscope. In a system of a mirror and multiple dielectric disks (booster) under a strong magnetic field $B_e$ each surface emits a coherent electromagnetic wave sourced by axions. Bringing those into constructive interference and making use of weak resonances in the system (not shown), the total amplitude can be much enhanced (characterized by the boost factor). The total emitted power can be received by focusing it onto an antenna. Adapted from~\cite{TheMADMAXWorkingGroup:2016hpc} with permission.
	}\label{fig:dielhaloscope}
\end{figure}
	The recently proposed dielectric haloscope~\cite{Jaeckel:2013eha,TheMADMAXWorkingGroup:2016hpc,millar2017dielectric} brings together both approaches by placing multiple dielectric disks in front of a metal mirror (booster). In this case each dielectric disk coherently emits electromagnetic radiation, which can interfere constructively within the setup and excite resonances between the disks including the mirror. These two effects give rise to the power boost factor $\beta^2$, which is defined in vacuum as the ratio between the power emitted by a dielectric haloscope\footnote{In this work we take antenna efficiency and reflection effects into account for this power.} and the power emitted by only the same mirror without dielectric disks. The boost factor is a function of frequency $\beta^2(\nu)$ and can be tuned to a specific frequency band by changing the distances between the disks.
	The signal from such a dielectric haloscope can then be steered by a focusing mirror into an antenna, cf.\ Fig.~\ref{fig:dielhaloscope}.
	
	The recently proposed MADMAX (MAgnetized Disk and Mirror Axion eXperiment) is a dielectric haloscope for the mass range between \SI{40}{\micro\electronvolt} and \SI{400}{\micro\electronvolt}, corresponding to \SI{10}{\giga\hertz} to \SI{100}{\giga\hertz}. It consists of up to $80$~lanthanum aluminate disks~\cite{TheMADMAXWorkingGroup:2016hpc,Brun:2019lyf}. Its power boost factor is designed to reach $\beta^2 \gtrsim \num{e4}$ over a bandwidth of $\sim \SI{50}{\mega\hertz}$. This leads to an emitted power of
	\begin{eqnarray}
	{P_\gamma}
	&=& 1.6\times 10^{-22}\,\si{\watt} ~ \left(\frac{\beta^2}{\num{5e4}}\right) 
	\left(\frac{A}{\SI{1}{\square\metre}}\right) \left(\frac{B_{\rm e}}{10~{\rm T}}\right)^2 \nonumber \\
	&& \times  \left(\frac{|C_{a\gamma}|}{1}\right)^2 \,  \left(\frac{\rho_a}{\SI{0.45}{\giga\electronvolt\per\centi\meter\cubed}}\right) ,
	\end{eqnarray}
	where $A$ is the disk area, $B_e$ is the external magnetic field parallel to the disk surfaces, $|C_{a\gamma}|$ is a model-dependent coupling constant proportional to the axion-photon coupling $g_{a\gamma}$ as defined in~\cite{millar2017dielectric}, and $\rho_a$ is the local dark matter density made up by axions. For the KSVZ~\cite{Kim:1979if,Shifman:1979if} and DFSZ~\cite{Dine:1981rt,Zhitnitsky:1980tq} benchmark axion models $|C_{a\gamma}|$ is $\sim 1.9$ and $\sim 0.7$, respectively.
	In order to cover a broad mass range, the disk positions of the final booster will be changed iteratively to scan different mass bands during a measurement procedure over several years. More detailed sensitivity estimates also in comparison with other experiments are presented in~\cite{Brun:2019lyf}.

	It is important to demonstrate that such a booster system with the desired electromagnetic properties can actually be realized.
	This is limited by the obtainable mechanical accuracy and systematic effects, such as unwanted reflections on the receiving antenna.
	In addition, it is essential to understand the implications of tuning the disk positions, for example by employing optimization schemes to match the predicted electromagnetic responses.  
	In this work we present a first proof of principle for such a booster setup. We evaluate systematic uncertainties and demonstrate basic tuning algorithms, as well as their impact on the power boost factor.
	
	To identify relevant observables we first briefly review the one-dimensional (1D) model of a dielectric haloscope, and then introduce the details of our experimental booster setup containing at present up to five sapphire disks and a copper mirror. 
	After discussing systematic limitations, 
	we describe the tuning of the setup and consequences of the systematic uncertainty on the power boost factor.

	\section{Electromagnetic Response Model}
	\label{sec:idealmodel}
		\begin{figure}
	    \centering
	    \includegraphics[width=\linewidth]{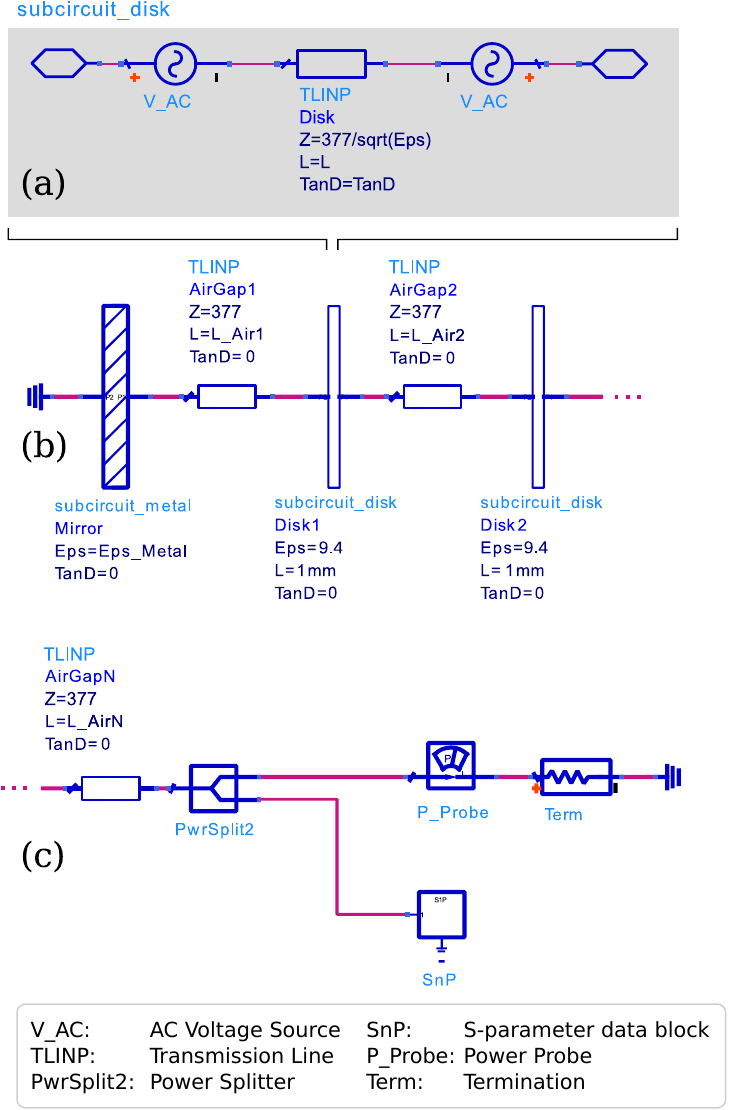}
	    \caption{One-dimensional (1D) model representation in \emph{Advanced Design System (ADS)}~\cite{ADS}. `Z' is the impedance in~\si{\ohm}, `Eps' is the dielectric constant, `L' is the length, and `TanD' is the loss tangent of the respective element. %
	    (a)~Model of a dielectric disk (`subcircuit\_disk'). The discontinuity of the axion-induced field on the disk surfaces is implemented with voltage sources (`V\_AC'). The copper mirror (`subcircuit\_metal') is modeled analogous. %
	    (b)~Booster model combining many dielectric disks. The mirror sub-circuit is analogous to the disk. %
	    (c)~Antenna model. For simplicity we use a power splitter diverting some of the power received by the antenna to the $S$-parameter data block (`SnP') where the measured antenna reflectivity can be included in the model. %
	    For details cf.~Sect.~\ref{sec:model:reflection}. %
	    }
	    \label{fig:model:ideal}
	\end{figure}
	
	\begin{figure}
	    \centering
	    \includegraphics[width=\linewidth]{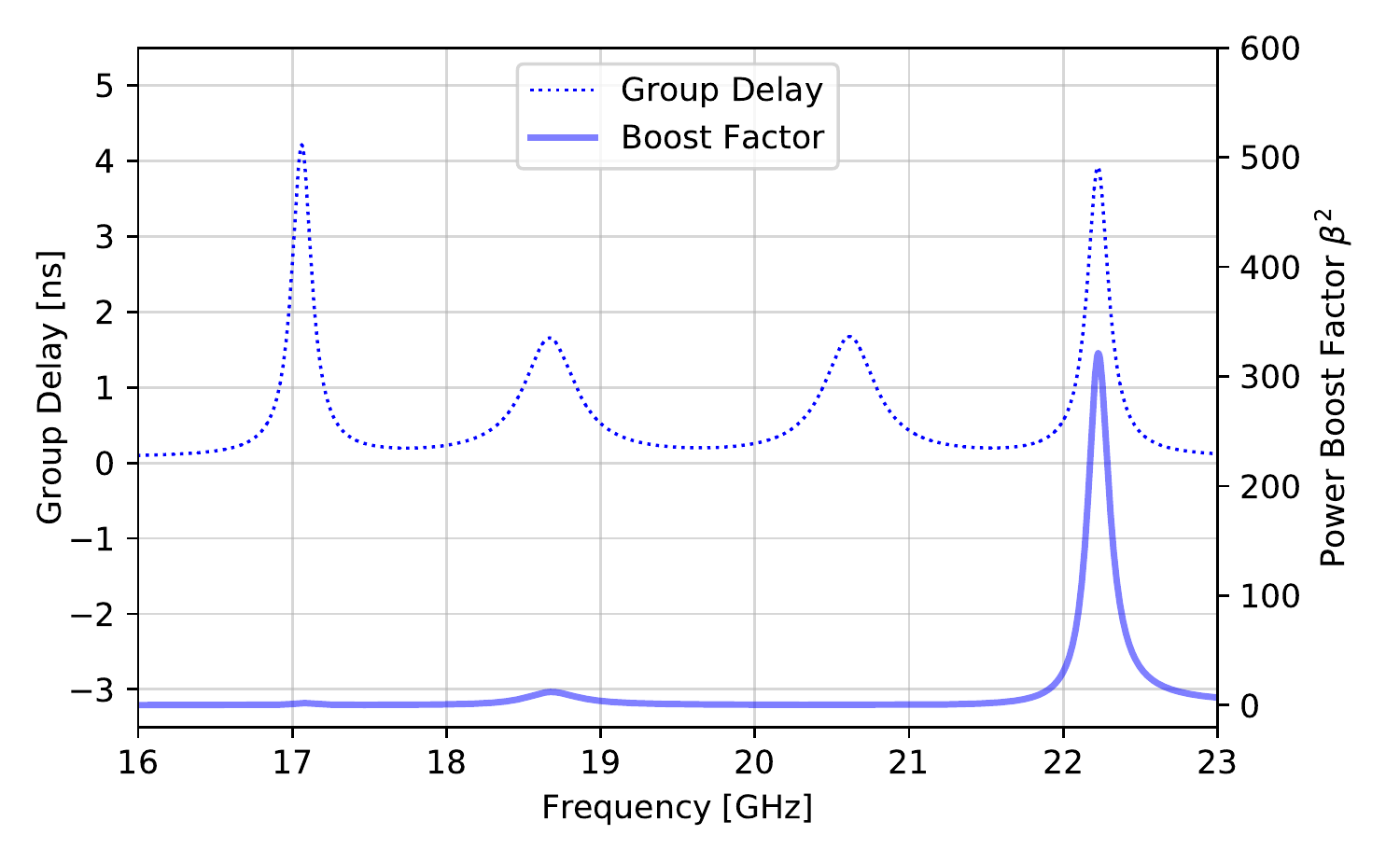}
	    \caption{Group delay and boost factor for a setup with 4 sapphire disks, each separated by \SI{8}{\milli\meter}, in the model with \SI{50}{\percent} antenna efficiency but without antenna reflections.
	    }
	    \label{fig:model:gd-bf}
	\end{figure}
	A detailed 1D model of such a booster using transfer matrices has been discussed in~\cite{millar2017dielectric}.
	Such 1D transfer matrices are implemented in circuit simulators such as \textit{Advanced Design System (ADS)}~\cite{ADS}.
	We show the circuit representing our proof of principle setup in Fig.~\ref{fig:model:ideal}. 
	The emission of electromagnetic radiation from the disk surfaces arises from the compensation of the discontinuity of an axion-induced electric field on the surfaces of the disks. Between two media this discontinuity is explicitly
	\begin{equation}
        \label{eq:Ea}
        \Delta {\bf E}_a(t)=-\left( \frac{1}{\epsilon_1} - \frac{1}{\epsilon_2} \right) g_{a\gamma} {\bf B}_{\rm e} \, a_0 \exp\left(-i \frac{m_a c^2}{\hbar}  t\right)\,,
    \end{equation}
	where $\epsilon_1, \epsilon_2$ are the dielectric constants of the media, $a_0$ is a normalized absolute amplitude of the axion-field, $c$ is the speed of light, $\hbar$ is the reduced Planck constant and $t$ is time.
	In the circuit simulator this discontinuity corresponds to a voltage drop between transmission-lines and can therefore be modeled using voltage sources at the disk surfaces.

	While the emitted radiation is caused by the axion field, 
	their propagation throughout the system follow the same physical laws as for an external test beam, which is reflected by the booster, i.e., an external signal can excite the same resonant modes inside the booster as the axion-induced signal.
	Therefore, the reflectivity $\Gamma$ of the haloscope is correlated with the boost factor $\beta^2$.  
	While the boost factor cannot be measured directly, the measured reflectivity can be directly compared with expectations from simulation to infer the boost factor and guide the tuning of the dielectric haloscope when changing the disk positions.
	In an ideal lossless booster the magnitude of the reflectivity will always be unity.
	Therefore, it is easier to use the phase $\Phi$ of the reflectivity rather than its absolute magnitude for frequency tuning.
	In practice we use the group delay $\tau_g = - d \Phi / d \omega$, where $\omega = 2\pi \nu$ is the angular frequency. 
    Qualitatively, the group delay can be understood as the mean retention time of reflected photons within the booster. Thus, peaks in the group delay frequency spectra map out resonances.	
    An example for the correlation between boost factor and group delay is explicitly shown in Fig.~\ref{fig:model:gd-bf} where we compare group delay and boost factor for a set of four equally spaced disks at $d = \SI{8}{\milli\metre}$ distances in front of a perfect mirror. 
	The group delay peak at $\sim \SI{22}{\giga\hertz}$ correlates with the one from the boost factor from the 1D calculation.
	The other peaks correspond to other resonant frequencies of the system where the emission from the individual disks, however, interferes destructively. This is analogous to the modes in resonant cavities that do not couple to the axion-induced field.

	\section{Experimental Setup}
	\label{sec:setup}
	\begin{figure}
	\centerline{
		\includegraphics[width=\linewidth,trim=450 300 950 150,clip]{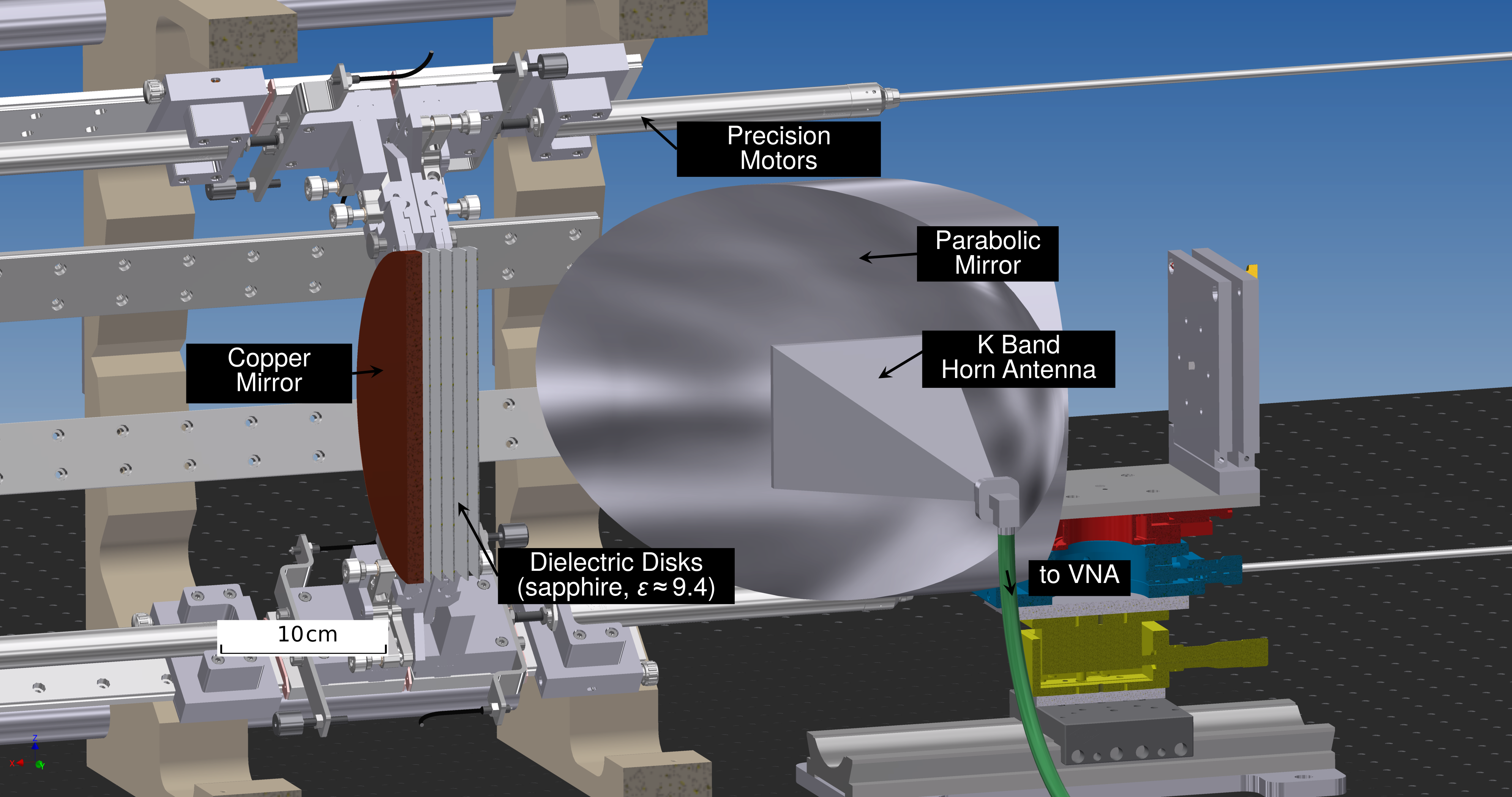}
	}
	\caption{
		Schematic drawing of the proof of principle setup. Up to 5 sapphire disks with a diameter of \SI{20}{\centi\metre} each are mounted in front of a plane copper mirror and can be moved using precision motors. Test beam reflections are studied via the parabolic-mirror-antenna assembly and measured with a vector network analyzer (VNA).
		For clarity we only show a cross section orthogonal to the disk surfaces.
		The distance between the booster and the antenna is adjusted to $\sim \SI{70}{\centi\metre}$ in our setup (shorter in this drawing for illustration purposes). The rest of the figure is to scale.
	}\label{fig:setup}
\end{figure}
We show a sketch of our experimental realization of such a booster in Fig.~\ref{fig:setup}. The proof of principle booster consists of up to five sapphire disks placed in front of a copper mirror.
The disks have a dielectric constant of ${\epsilon \approx 9.4}$ perpendicular to the beam axis (crystal C-axis), 
a thickness of $\SI{1}{\milli\metre}\pm \SI{10}{\micro\metre}$,
and a diameter of \SI{20}{\centi\metre}.
Each disk is mounted on a holder connected to its own DC motor (\emph{PI~L-220.70DG}~\cite{PIMotor}). The motors are equipped with a position encoder and change the disk positions with a precision of less than a ${\mu \rm m}$.
The tilt of the disks are adjusted at the \si{\milli\radian} level by using tilting stages on the disk holders. The tilts are manually optimized to maximize the measured absolute reflectivity of the booster.
The disk holders slide on rails which have been aligned such that they are parallel up to $\sim \SI{100}{\micro\metre}$ over a length of $\sim \SI{0.5}{\metre}$~\cite{Partsch:2018}.
To determine the temperature of the setup, Pt-100 sensors are placed at the metal frame, on the backside of the copper mirror, and in the air above the disks.

To determine the electromagnetic response of this booster we measured the phase and amplitude of a reflected microwave signal in the frequency range of ${10 - \SI{30}{\giga\hertz}}$ using a Vector Network Analyzer (VNA). 
It is connected to a rectangular horn antenna (\textit{A-INFO LB-42-25}~\cite{Antenna}) facing a 90-degree off-axis parabolic aluminum mirror with a diameter of $\sim \SI{30}{cm}$.
The effective focal length of this mirror is $\sim \SI{15}{\centi\metre}$, such that the focal point lies within a few \si{\centi\metre} at the antenna waist.

By modifying the tilt and position of the parabolic mirror we centered the resulting beam shape on the disks and minimized its incident angle.
We adjusted the alignment of the beam without sapphire disks installed. 
To this end, we distorted the beam from different transverse sides with a rectangular metal reflector. If the beam is symmetric and centered, this leads to approximately the same change in the reflected signal.
The accuracy of this method is better than $\sim \SI{1}{\centi\metre}$ for the transverse position of the beam.

The reference plane of the VNA has been calibrated to the connector end of the horn antenna. By taking a reflectivity measurement without any dielectric disk installed, we measured the transmission coefficient between the coax connector end and the copper disk, including diffraction losses and the transmission factors of the antenna-parabolic-mirror assembly.
Assuming a lossless copper mirror, we obtained the reflectivity of a booster, i.e., a system including disks, by dividing its measured reflectivity with this factor.
	
	\section{Systematic Limitations}
	\label{sec:systematics}
	    The precision at which this system can be adjusted to a certain boost factor is limited by the mechanical precision of the setup and systematic microwave effects, such as unwanted reflection and losses. In this section we survey these effects.
    
    \subsection{Mechanical Stability}
    
    \begin{figure}
        \centering
        \includegraphics[width=\linewidth]{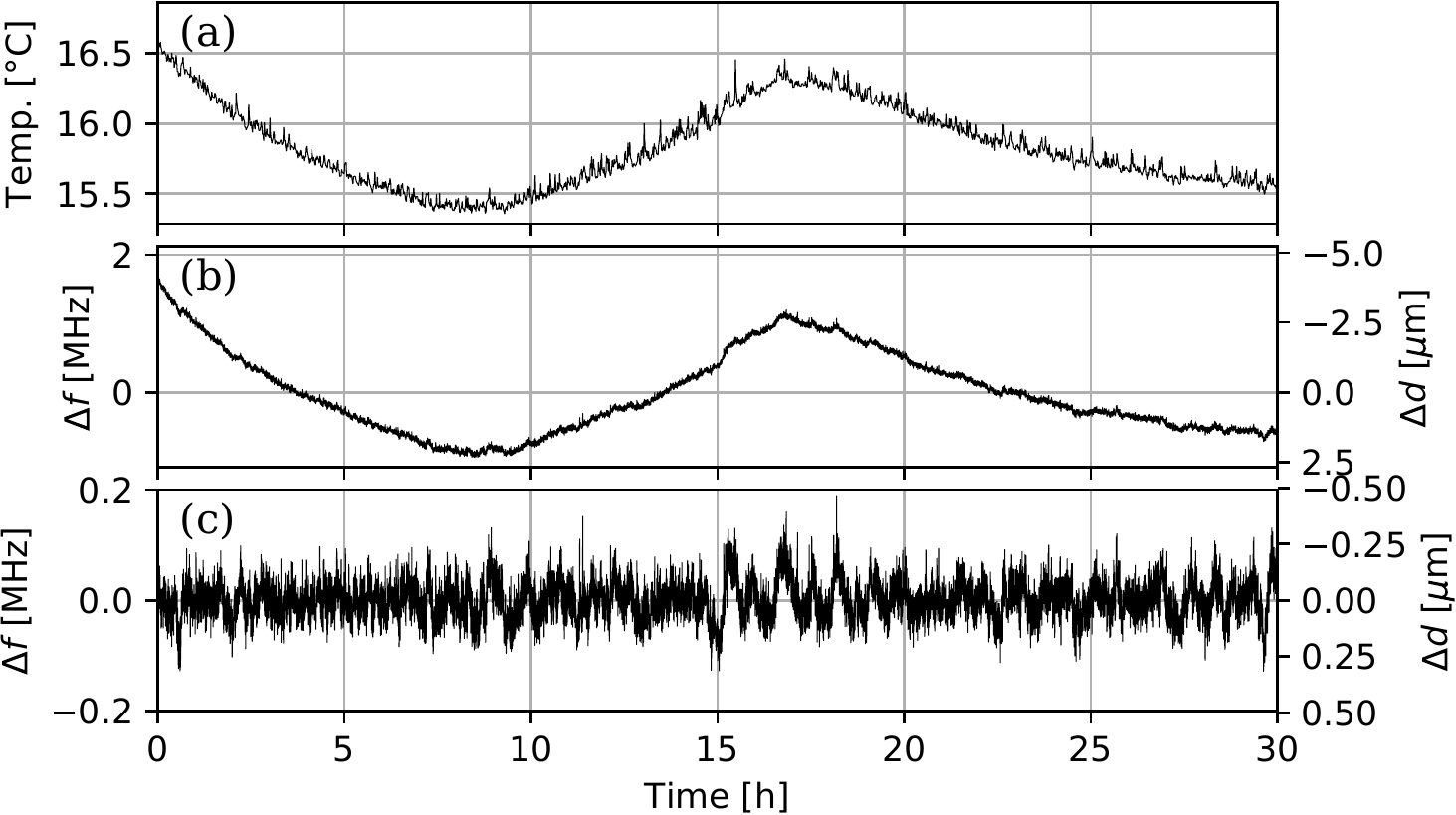}
        \caption{Effect of mechanical deformation of the proof of principle setup over a long-term measurement with a single disk and mirror. (a)~Measured temperature as a function of time. (b)~Measured frequency of the 7th harmonic of the group delay peak as a function of time. The secondary axis shows the corresponding positioning error of the dielectric disk. More details are given in the text. (c)~Measured group delay peak frequency and corresponding position deviation after subtracting variations on the time scale of an~hour.}
        \label{fig:mechanical_stability}
    \end{figure}

    The mechanical stability of the system is limited by vibrations from the surrounding, as well as long term displacement effects during measurement caused, for example, by temperature changes. 

    We investigated the long-term mechanical stability of our proof of principle setup by observing a resonant group delay peak of a single disk and mirror system.
    Fig.~\ref{fig:mechanical_stability} shows the result of a long-term measurement where the sapphire disk was placed $\sim \SI{5}{\centi\metre}$ apart from the mirror without any motor movements. (a) shows the mean of all measured temperatures as a function of measurement time. (b) shows the frequency position of the group delay peak of the 7th harmonic from the resonance between the sapphire disk and copper mirror at $\sim \SI{21}{\giga\hertz}$. (c) shows the short-term fluctuations in (b).
    We obtained the residual short-term fluctuation by subtracting a smoothed version of (b), which has been obtained by applying a Gaussian filter with one hour standard deviation.
    The frequency stability is mainly limited by temperature effects of  $\sim\,\SI{2}{\mega\hertz \per \kelvin}$, corresponding to a dependency of the disk position of \SI[separate-uncertainty = true]{-5(3)}{\micro \meter \per \kelvin}~\cite{Egge:2018} consistent with the expected thermal expansion of the motor rods. This value slightly varies with motor and motor position, 
    thus the systematic uncertainty is conservative.
    Frequency analysis of the displacements over time reveal that the residual variations can be attributed to vibrations such as table movements.
    They leave the group delay peak stable well below the \si{\mega\hertz} scale.
    The accuracy of the disk alignment is limited by the precision of the motors and the mechanics of the disk holders. 
    We determined the reproducibility of positioning individual disks by repeatedly placing a single sapphire disk to a predefined position now $\sim \SI{8}{\milli\metre}$ apart from the mirror.
    The measurement revealed a mechanical hysteresis on the order of $\sim \SI{2}{\micro\metre}$ positioning difference depending whether the motor is placed from a higher or lower position. When mitigating this effect by placing the disk always from the same side, the group delay peak position was reproduced within a standard deviation of $\sim \SI{1.5}{\mega\hertz}$. This corresponds to position errors of the order of $\sim \SI{0.7}{\micro\metre}$ in this case.
    This is still well within the needed accuracy for a boost factor bandwidth of \SI{250}{\mega\hertz} even for 20 and 80 disk systems where both $\mathcal{O}(\SI{10}{\micro\meter})$ is required,~cf.~\cite{millar2017dielectric,Brun:2019lyf}.

	\subsection{Unwanted Reflections}
	\label{sec:model:reflection}
	\begin{figure}
	    \centering
	    \includegraphics[width=1.0\linewidth]{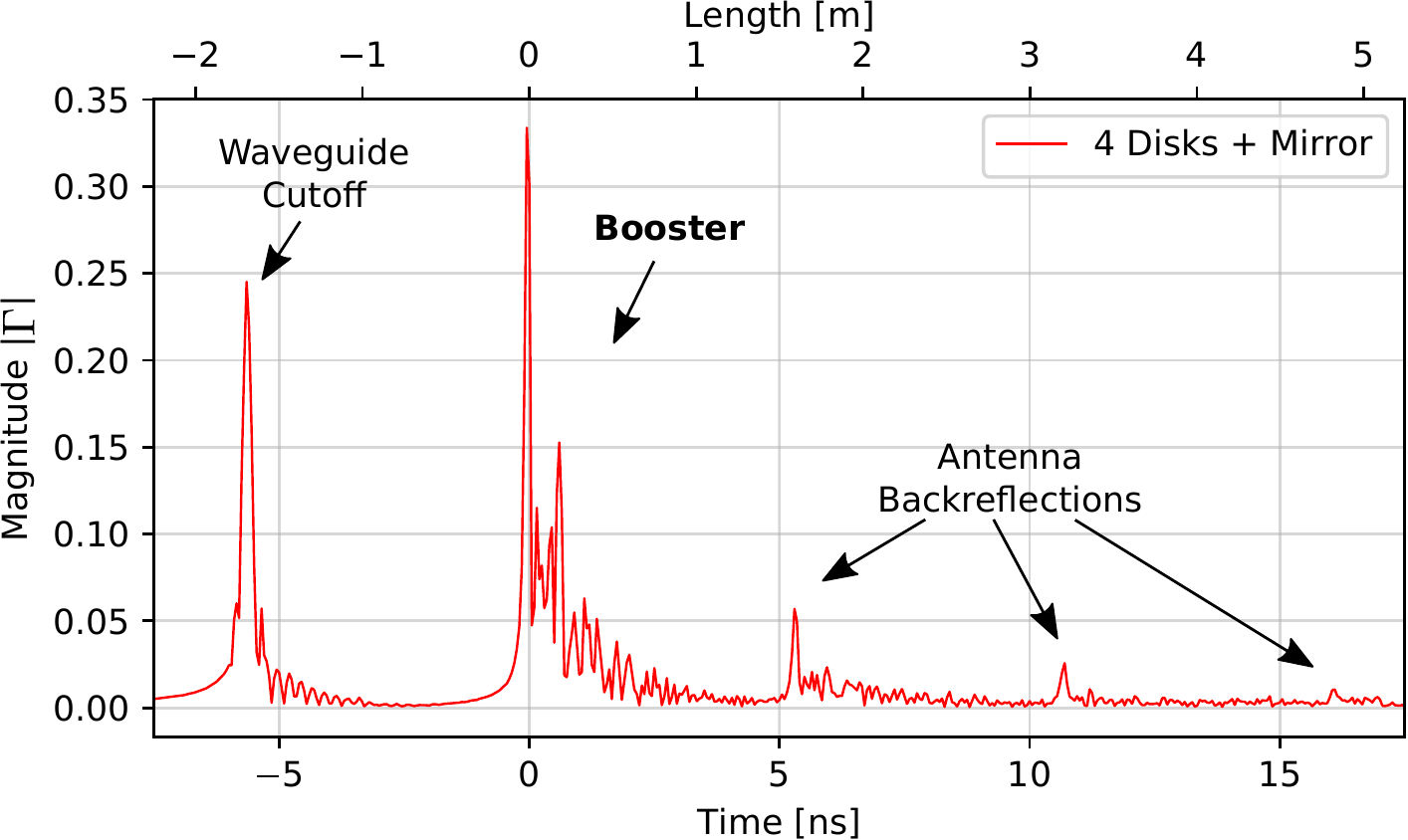}
	    \caption{Measured reflectivities in time domain for a setup with the copper mirror and 4 sapphire disks. The corresponding length assumes free space propagation.}
	    \label{fig:model:time-domain}
	\end{figure}
	An important systematic effect on the measurement of the electromagnetic response results from unwanted reflections on the antenna.
	Fig.~\ref{fig:model:time-domain} shows the time domain responses for a setup with four disks in front of the mirror. The booster reflectivity is essentially described by an exponential decay in time domain corresponding to its resonant behaviour. The time scale for the exponential decay increases when making the system more resonant, e.g. by adding more dielectric disks. The time domain data is shifted such that the reflection on the front-most disk of the booster happens at $t_0 = 0$.
	The other peaks can be identified according to the time they occur. The time between two such peaks is $\sim \SI{6}{\nano\second}$. This roughly corresponds to twice the distance between the antenna-coax-connector and the booster of $\sim \SI{90}{\centi\metre}$ and increases accordingly, if the booster is moved farther away from the antenna. The peak at $\sim \SI{-6}{\nano\second}$ corresponds to an internal reflection inside the antenna back to the VNA, mainly arising from the cutoff frequency of the waveguide section at around $\sim \SI{14}{\giga\hertz}$. The peak at $\sim \SI{6}{\nano\second}$ corresponds to the reflection off the antenna back to the booster. This reflection repeats every other $\sim \SI{6}{\nano\second}$ since a reflected signal may be back-reflected once again.
	For a low number of disks we can filter out this higher order reflection in time domain by simply multiplying with a window function (time gating). For more resonant systems the main peak stretches over a longer time period and begins to overlap with the unwanted reflection.
	In addition, the axion-induced signal may also be affected by the reflection. However, time gating can only be done with a reflected signal, which can be related to the phase of the input signal over a broad frequency range. The axion signal cannot be time gated since it is continuous in time, oscillates in a narrow frequency band within a few \si{\kilo\hertz} and its phase is unknown.
	
\begin{figure}
	    \centering
	    \includegraphics[width=0.5\textwidth]{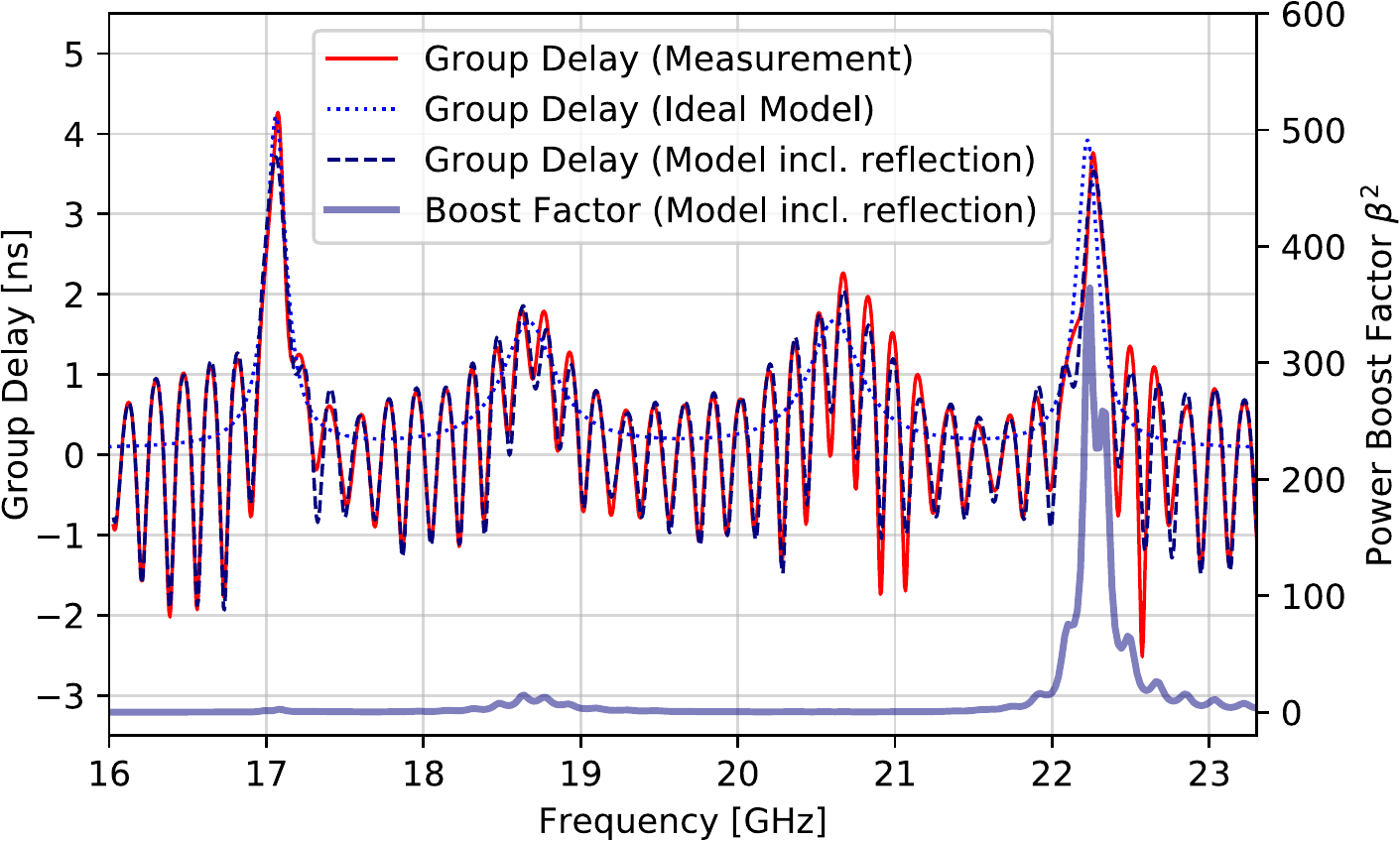}
	    \caption{Group delay including antenna reflection from measurement (red straight line), without antenna reflection in idealized loss-less model (blue dotted line) and after including the antenna reflection (darkblue dashed line), as well as the corresponding boost factor including the antenna reflection (bold lightlblue line). For the same 4 disk booster as in Fig.~\ref{fig:model:gd-bf}. We only show a subsection from all recorded frequencies containing the resonant group delay peaks of the booster.}
	    \label{fig:model:gd-incl-reflection}
	\end{figure}

	Fig.~\ref{fig:model:gd-incl-reflection} shows the effect of the first order reflection on the group delay and the boost factor as opposed to the ideal case shown in Fig.~\ref{fig:model:gd-bf}.
	The longer runtime of the reflected signal compared to the main signal causes a phase difference between them, which changes linearly with frequency. Depending on frequency the interference is constructive or destructive, which adds an additional sinusoidal structure (ripples) to the frequency spectra of the group delay and the boost factor.
	This changes the shape of the group delay peaks. 
    When fitting the group delay of an ideal 1D model with the disk positions as free parameters to such data, 
    the predicted boost factor would therefore end up at the wrong frequency.
	The distance between the antenna and the booster defines the frequency difference between neighbouring ripples and thus the maximum frequency shift this effect may cause. For our distance of around $\sim \SI{90}{\centi\metre}$ this shift is at the order of $\sim \SI{200}{\mega\hertz}$.
	
	For these reasons, we minimized the reflections in the setup by screening reflective parts of the antenna with crumpled copper foil and maximizing the diameter of the parabolic mirror. This could reduce some of the reflections by up to a factor $\sim 4$.
	The residual reflection coefficient of the antenna ${\Gamma}_A$ was separately measured and included in the electromagnetic model, which we compare later with our measurements.
	${\Gamma}_A$ can be obtained measuring the reflectivity only with the copper mirror and without dielectric disks installed. By applying a gate in time domain around the peak of the antenna reflection at $\sim \SI{6}{\nano\second}$ to the measured reflectivity, we can infer the antenna reflectivity assuming the copper mirror is lossless.
	This reflectivity can be directly included into the 1D model afterwards. 
    For simplicity we only include the reflectivity as seen from outside the antenna rather than the reflectivity from inside and transmission factors as in a full antenna model. We do this by adding a power splitter\footnote{We set the transmission factors $S_{21} = S_{31} = 1/\sqrt{2}$, which leads to an antenna efficiency of \SI{50}{\percent} in the model.} that diverts some of the reflected power to a reflective element.
	Its reflectivity is set such that the total reflectivity of our antenna model in Fig.~\ref{fig:model:ideal}\,(c) is $\Gamma_A$.

	We tested these improvements by measuring the group delay of the same booster, but at varying distance from the antenna over more than the range of a full wavelength. Afterwards, we fit the data with the model including the unwanted reflections.
	We find that when changing the distance the boost factor maximum frequency is shifted by less than \SI{20}{\mega\hertz}~\cite{Egge:2018}.
	This is still much smaller than the full-width-half-maximum (FWHM) of, for example, the boost factors shown in Fig.~\ref{fig:model:gd-bf} and Fig.~\ref{fig:model:gd-incl-reflection}, which have a FWHM of $\sim \SI{150}{\mega\hertz}$.
	Notice that this limitation can be in principle further improved by placing the antenna farther away from the booster, since this decreases the distance between the frequencies of the ripples on the group delay.
	
	\subsection{Losses and Other Effects}
	
	There are various ways electromagnetic power can dissipate in a dielectric haloscope, including dielectric loss, diffraction, tilts, and surface roughness of the disks. For small losses we can parametrize those using the usual loss tangent $\tan \delta$ in the 1D model.
	To first order these reduce the total power in the system, i.e., the power boost factor and the absolute reflectivity.
	Therefore, the absolute reflectivity may be used in a later stage to calibrate the losses in the system and predict their impact on the power boost factor.
	This will in particular lead to a more realistic estimate of the boost factor amplitude.
	However, in addition losses affect the phase of a propagating electromagnetic wave. Therefore, they can also cause changes in the group delay and even lead to negative group delay peaks~\cite{Takahashi:2004,Das:2017:synthesis}. 
	
	By minimizing losses from tilts and only considering a system with up to five sapphire disks, we ensured the above losses are small enough\footnote{Corresponding to $\tan \delta \lesssim 10^{-2}$ in the 1D model.} such that the lossless model can still approximately reproduce the measured group delays. In the following we will not consider the systematic effects caused by various loss mechanisms and focus on the task to tune the booster, i.e., we will compare our measurements to an ideal lossless model.
    While this does not take into account systematic effects in 3D and from losses, it is sufficient to evaluate the alignment reproducibility of the real setup. It also lets us assess to what extend the behaviour of the electromagnetic response as a function of disk spacings is as expected from the idealized model.
    First results on the impact of diffraction losses on dielectric haloscopes from 3D simulations are available~\cite{Knirck:2019eug} and more detailed studies are~\mbox{underway}.

	\section{Tuning}
	\label{sec:optimization}
	Since different axion masses correspond to different frequencies, the boost factor needs to be tuned over the frequency range of interest during the axion search. We demonstrate such tuning can be achieved with up to five equidistant disks using the following tuning procedure: First, the disk positions corresponding to the desired boost factor are calculated with the 1D model, along with the corresponding group delay of the reflectivity. Then, the disk positions in the setup are adjusted by an optimization algorithm in order for the group delay to match the prediction. In the last step we test this tuning procedure by fitting the model to the measured data to infer the systematic uncertainty on the boost factor arising from tuning.

\begin{figure}
    \centering
    \includegraphics[width=\linewidth]{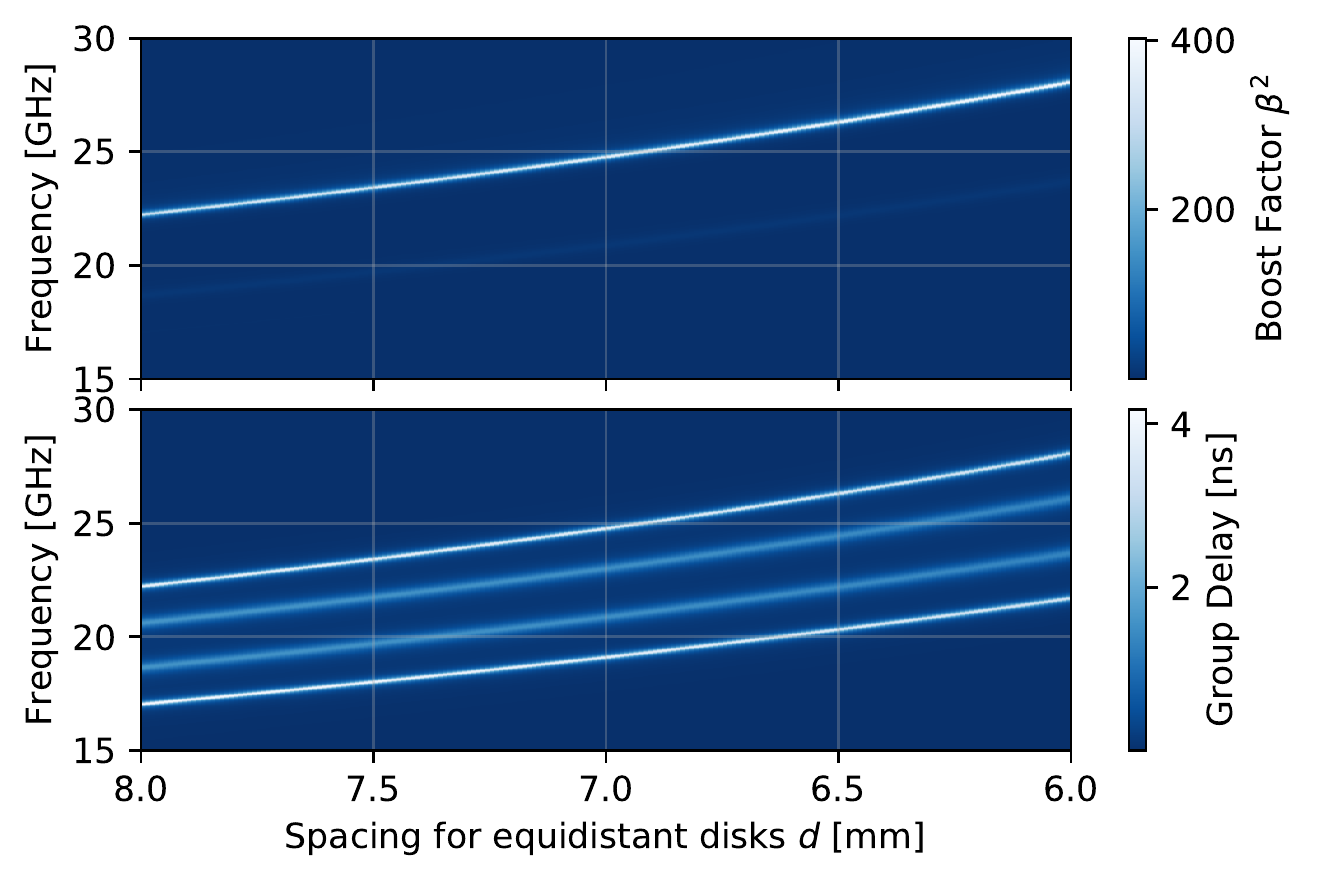}
    \caption{Simulated boost factor (top) and group delay (bottom) as a function of frequency (vertical axes) and separation between the sapphire disks (horizontal axes) for a system of four equidistant disks.}
    \label{fig:bf_tuning}
\end{figure}

In the first step, we adjust the boost factor to the right frequency band only within the model. This can be done numerically by optimizing the disk spacings in the simulation such that the desired boost factor is obtained.
For simplicity here we do not optimize the disk positions, but consider configurations where all disks are at the same distance $d$. Fig.~\ref{fig:bf_tuning} shows how the boost factor and group delay change for a four disk booster as a function of $d$. By tuning $d$ between \SI{6}{\milli\metre} and \SI{8}{\milli\metre} the boost factor can be tuned over a frequency range between \SI{22}{\giga\hertz}~to~\SI{28}{\giga\hertz}.

\begin{figure}
    \centering
    \includegraphics[width=\linewidth]{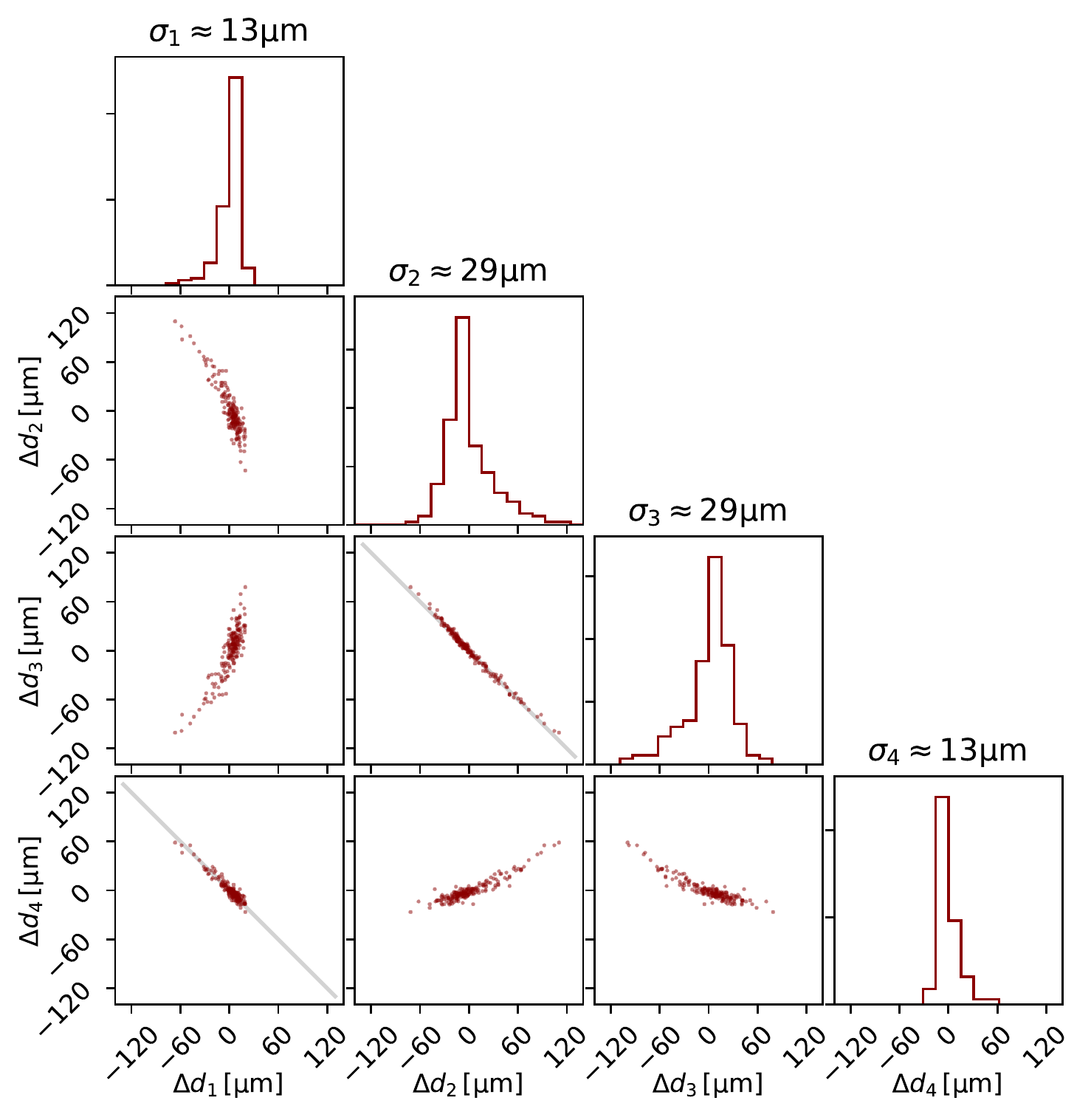}
    \caption{Final disk spacings after optimizing the actual physical disk positions for the four disk setup for $\sim 200$ times. $d_1$ refers to the spacing between the copper mirror and the first disk, $d_2$ the following spacing, and so on. $\Delta d_i$ is the difference of the final spacing $i$ compared to its mean. The diagonal subplots show the histogram of final disk spacings, the off-diagonal subplots show scatter plots illustrating the correlation between different gaps. The gray lines indicate $\Delta d_2 = - \Delta d_3$ and $\Delta d_1 = - \Delta d_4$.}
    \label{fig:corellation}
\end{figure}
In the second step we aim for aligning the real disk positions in the experimental setup such that it produces the simulated boost factor.
However, simply placing the disks at the distances predicted from the 1D model and relying on the simulated boost factor is not the most feasible solution. Effects such as disk bending, surface roughness, or variations in the dielectric constant may change the phase depths of each region and therefore contribute to the systematic uncertainty on the boost factor.
Thus, it is more straightforward to optimize the electromagnetic properties of the system directly.
Since the boost factor itself cannot be measured, we therefore align the disk positions such that the measured group delay matches the simulated group delay. It is related to the boost factor as described in Sect.~\ref{sec:idealmodel}.

The error function for such an optimization is the sum of the squared differences between the measured and desired group delay data points as a function of frequency. Each evaluation corresponds to a realignment of disk positions and a measurement of the reflectivity. Therefore, we use gradient-free optimization algorithms that minimize the number of evaluations, while trying to be robust against local minima. 
Various optimization algorithms~\cite{Whitley1994,Laarhoven:1987:SAT:59580,Lagarias1998} lead to consistent results, with small differences observed in the number of error function calls and convergence. 
As a convergence criterion we require changes of the motor positions of less than a $\si{\micro\metre}$ between subsequent iterations.

\begin{figure}
    \centering
    \includegraphics[width=\linewidth]{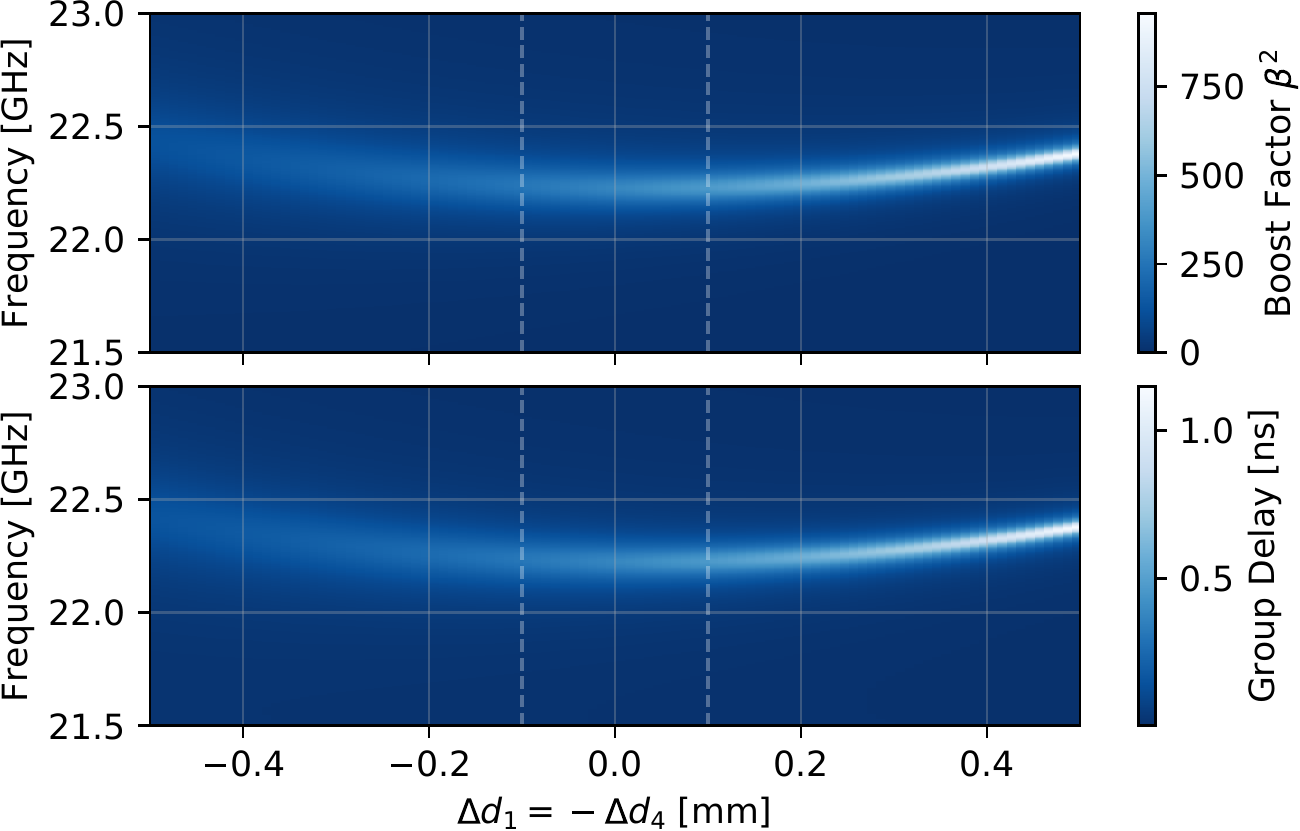}
    \caption{Simulated boost factor (top) and group delay (bottom) as a function of frequency (vertical axes) and separation between the mirror and first disk $\Delta d_1$ and changing the separation between the third and fourth disk as $\Delta d_4 = -\Delta d_1$, i.e., following the gray line in the leftmost subplot in the bottom row of Fig.~\ref{fig:corellation}. The vertical dashed lines indicate the $\pm \SI{100}{\micro\metre}$ range comparable to the range of the correlations in Fig.~\ref{fig:corellation}. }
    \label{fig:bf_follow_correlation}
\end{figure}
We evaluate the reproducibility of the system response by repeatedly performing such an algorithm using uniformly distributed randomized starting positions within $\SI{100}{\micro\metre}$. 
Fig.~\ref{fig:corellation} shows the differences in the final motor positions when running the frequency tuning procedure for $\sim 200$ times using the same calculated boost factor function with four disks and \SI{8}{\milli\metre} disk spacings.
The correlations show that there exists degeneracy in the disk position phase space.
This degeneracy can be understood by observing that the group delay and boost factor maxima frequencies are at first order unchanged when changing the disk spacings along the correlation, as explicitly demonstrated in Fig.~\ref{fig:bf_follow_correlation} for the correlation between the first and last spacing in the four disk setup --- as opposed to, for example, the changes used for tuning outlined in~Fig.~\ref{fig:bf_tuning}.

\begin{figure}
    \centering
    \includegraphics[width=0.6\linewidth]{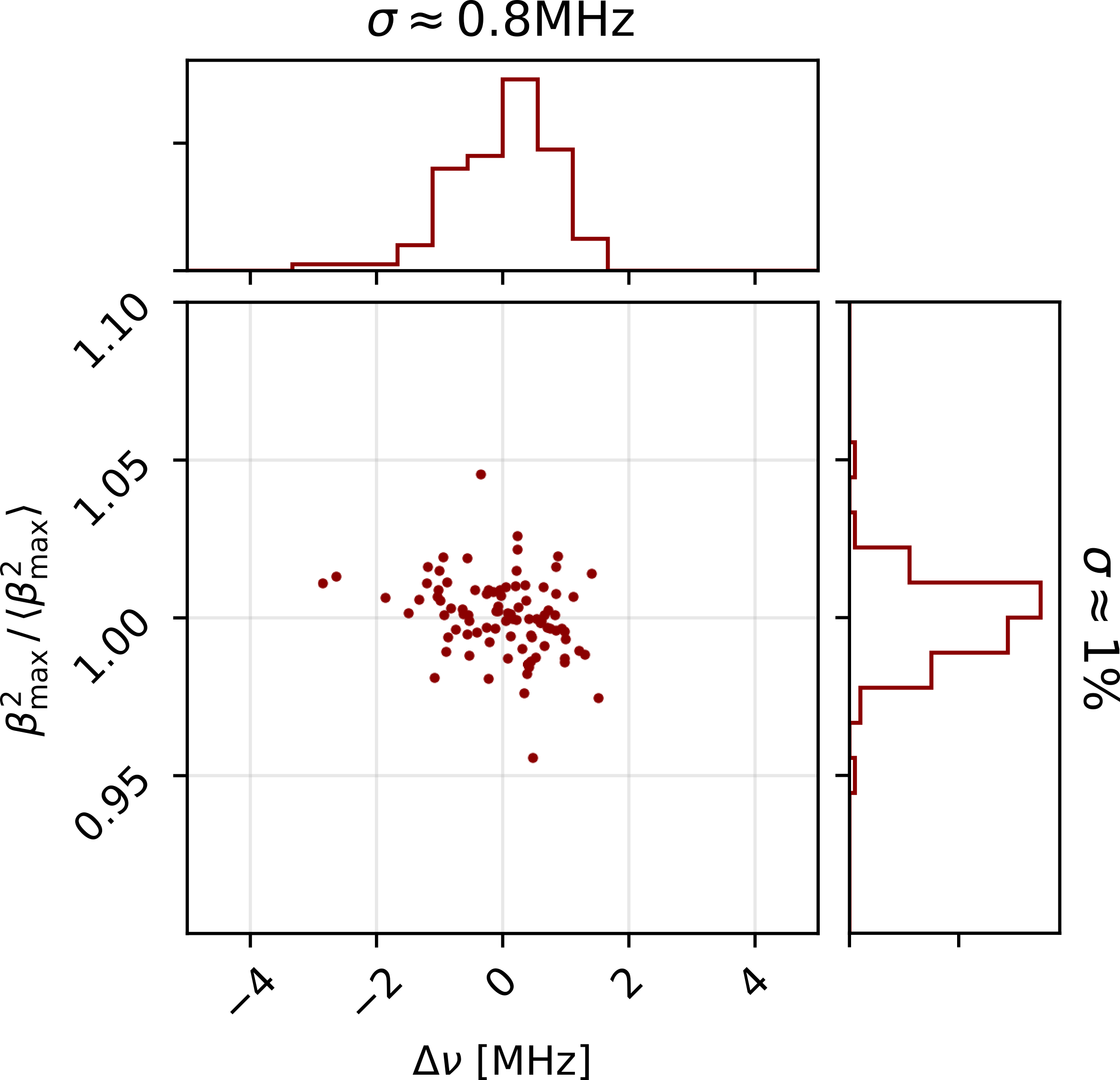}
    \caption{Scatter plot of the obtained power boost factor frequency positions and amplitudes after fitting the model to the measured group delays corresponding to the physical disk positions shown in Fig.~\ref{fig:corellation}. The horizontal axis shows the frequency shift of the boost factor maximum frequency compared to the mean of the considered ensemble. The vertical axis shows the relative boost factor amplitude deviation compared to the mean of the ensemble.}
    \label{fig:bf_err_4disk_only}
\end{figure}
To quantify the systematic uncertainty on the boost factor arising from this tuning procedure,
we optimize the disk positions in the 1D model until the calculated group delays best match the measured ones.
Afterwards we calculate for each realization the corresponding boost factor. The systematic uncertainty on the obtained maximum of the boost factor can then be extracted from this ensemble, thus taking into account the effect of disk spacing degeneracy.
The degeneracy in disk position parameter space is reproduced for the simulated disk positions. This confirms that the effect is explained by a degeneracy of the electromagnetic response expected from the model.
Fig.~\ref{fig:bf_err_4disk_only} shows a scatter plot of the obtained relative boost factor amplitudes against the shift of the maximum boost factor frequency for the same 4 disk case as in Fig.~\ref{fig:corellation}. The disk spacing degeneracy does not lead to a degeneracy in the boost factor frequency and amplitude phase space.
\begin{figure}
    \centering
    \includegraphics[width=\linewidth]{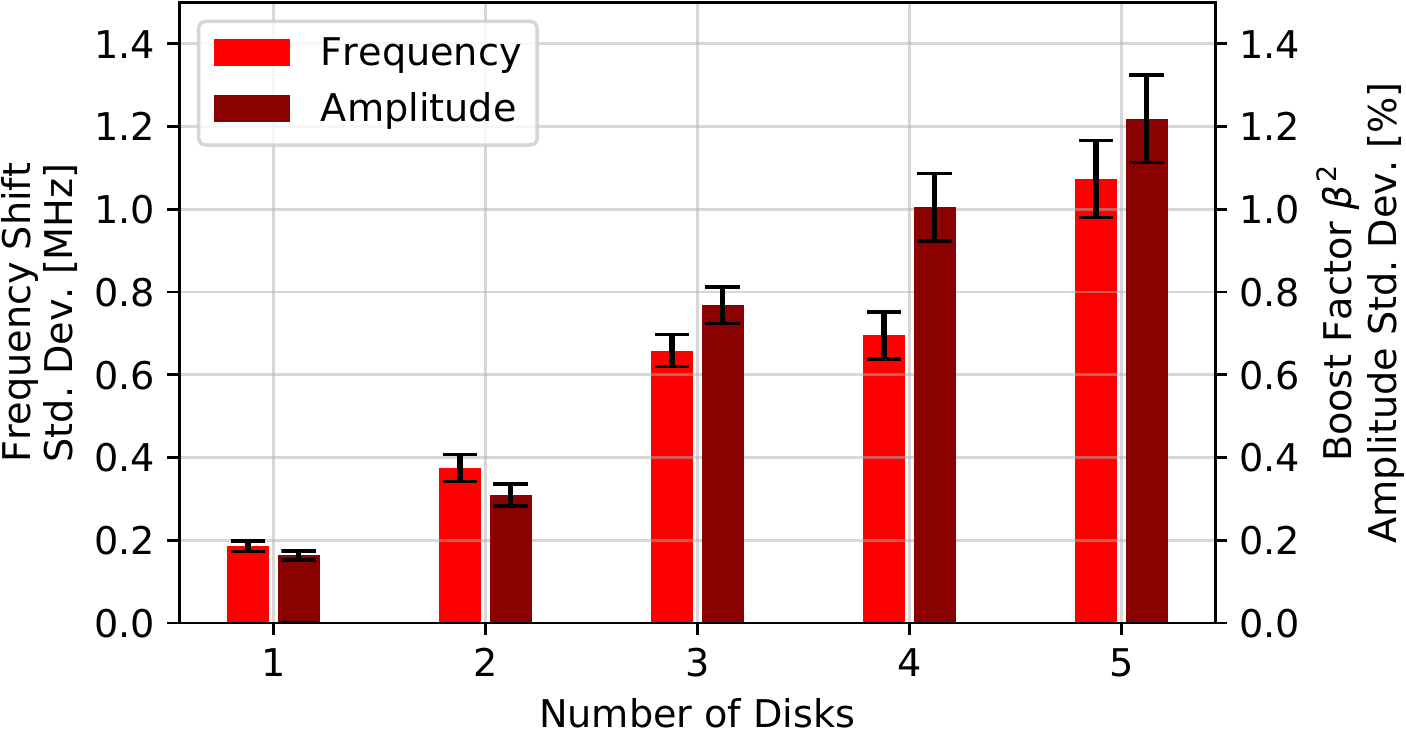}
    \caption{Standard deviation on frequency position (left bars, light red) and standard deviation on amplitude relative to the ideal amplitude (right bars, dark red) of the boost factor after fitting the model to the measured group delays corresponding to the physical disk positions for one to five disks. More details cf. text.}
    \label{fig:bf_err}
\end{figure}
Fig.~\ref{fig:bf_err} shows the resulting uncertainties as a function of the number of disks for a measurement for setups resulting in a boost around 22\,GHz like in Fig.~\ref{fig:model:gd-incl-reflection}. For this case, the frequency uncertainty ($< \SI{2}{\mega\hertz}$) and the amplitude uncertainty ($< \SI{2}{\percent}$) would have a small impact on the sensitivity of the booster. Note that for a fixed disk number the width of the boost factor peak is approximately inversely proportional to the boost factor amplitude according to the area law in~\cite{millar2017dielectric}, i.e., the relative uncertainty on the FWHM is approximately the same as the relative uncertainty on the boost factor amplitude.
This result does not contain possible systematic uncertainties not covered by the lossless 1D model.
Analogous results have been obtained for other equidistant configurations in the frequency range between \SI{22}{\giga\hertz}~to~\SI{28}{\giga\hertz}.

The correlations in disk spacings and the resulting boost factors illustrate that different disk spacing configurations can still result in the same electromagnetic response. This shows that phase errors can be compensated by corresponding phase errors resulting from other disk spacings.

	\section{Conclusions}
	\label{sec:conclusion}
	In this work we have presented a first tunable proof of principle booster for the MADMAX dielectric haloscope. It consists of a copper mirror and up to five sapphire disks of a diameter of \SI{20}{\centi\metre} and is tunable over the frequency range between \SI{22}{\giga\hertz} and \SI{28}{\giga\hertz}.
In order to derive the boost factor of the system, the correlation with the group delay of a reflected signal is used.

Using our setup, we have surveyed various systematic effects which have adverse impact on the reflectivity measurement and boost factor. 
The mechanical stability of the system corresponds to a frequency stability at the order of \si{\mega\hertz}, which is sufficient for the concept of broadband dielectric haloscopes.
A major limitation for the measurement of the group delay is placed by unwanted reflections from the antenna. By employing strategies to reduce those and calibrate them within the model, we have shown how this can be reduced to less than \SI{20}{\mega\hertz} in our setup, which limits the minimal usable boost factor bandwidth. By employing an antenna with a smaller reflection coefficient and increasing the distance to the booster this effect can be further reduced.

Moreover, we have demonstrated that our system can be tuned to achieve a preset desired electromagnetic response. The measured group delays follow the expected dependencies on disk spacing predicted by the model. In particular, we see degeneracy in the disk spacing phase space, which is explained by the 1D model. It allows for compensating misaligned disks on one side with slightly offset spacings on the other side. Although the physical disk positions ended up in a range over more than \SI{100}{\micro\metre} after our tuning procedure, the measured electromagnetic response always corresponded to a boost factor with its amplitude remaining unchanged up to a few percent and which was shifted in frequency by less than \SI{2}{\mega\hertz}. While this shows that the repeatability of the tuning procedure is sufficient, this does only include systematic effects covered by the lossless model, i.e., no losses and 3D effects currently under study~\cite{Knirck:2019eug}.
The degeneracy shows that phase errors, arising for example from position errors or surface imperfections of the disks, can be compensated by corresponding phase errors resulting from other disks.

In summary, we have demonstrated that a stable, small-scale dielectric haloscope with the predicted electromagnetic properties from a 1D model can in fact be realized and achieves the necessary accuracies, which is an important milestone for the realization of MADMAX.
The setup is currently being extended with an independent disk position measurement, a better antenna and higher disk numbers, which will be the next step before going to the MADMAX prototype with 20 disks with a diameter of \SI{30}{\centi\metre}.
	
	\begin{acknowledgments}
	The authors would like to thank Allen Caldwell, Chang Lee, Xiaoyue Li, Javier Redondo, Derek Strom and the MADMAX collaboration for interesting discussions and helpful comments.
	We thank Moahan Murugappan for his assistance with optimization algorithms.
	We thank Chris Gooch, Armen Hambarzumjan, David Kittlinger, Alexander Sedlak and G\"unther Winkelm\"uller for technical assistance.
	This work was funded by the Max-Planck society (MPG) and partly by the German Research Foundation (DFG) via the Excellence Cluster Universe and the Collaborative Research Center, SFB 1258.
	\end{acknowledgments}

\bibliographystyle{unsrtnat}
\bibliography{bibliography}

\end{document}